\documentclass[11pt,a4paper]{article}

\begin{document}
\title{It From Bit and The Unsmooth Reality
       \thanks{submitted to FQXi 2013 Essay Contest - ``It from Bit or Bit from It''} }
\author{ Joachim J. W{\l}odarz \\
         Dept. of Theoretical Chemistry and Computer Science Group \\
         Faculty of Mathematics, Physics and Chemistry \\
         University of Silesia, Poland \\
         \texttt{jjw@us.edu.pl}
       }
\date{June 28, 2013}
\maketitle
%%\bigskip
\abstract{In this short essay it is argued that the ``It from Bit'' idea is plausible when
          assuming ``generalized bits'', resulting from the Kolmogorov superposition 
          theorem, as universal building blocks. }

\bigskip
\bigskip

It from Bit or Bit from It ?  Is Reality Digital or Analog ?  These two questions,
theming the present and one of the past FQXI contests are deeply interrelated and
still controversial.

{\em It} is usually assumed to be an "element of reality", something we can isolate
or prepare and investigate, at least in principle. On the other hand, {\em Bits} are 
"elements of information", something we can associate with a sequence of {\em Yes}/{\em No} 
or {\em True}/{\em False} statements, the simplest answers we may get from an investigation. 
Therefore, as put sometime ago by Wheeler \cite{JAW_1998}

\begin{quotation}
[...] it is not unreasonable to imagine that information sits at the core of physics, 
just as it sits at the core of a computer.
\end{quotation}

Wheeler himself was deeply convinced that it is information what really matters 
\cite{JAW_1990}: 

\begin{quotation}
[...] every "it" — every particle, every field of force, even the space-time continuum 
itself — derives its function, its meaning, its very existence entirely — even if in 
some contexts indirectly — from the apparatus-elicited answers to yes-or-no questions, 
binary choices, bits. "It from bit" symbolizes the idea that every item of the physical 
world has at bottom — a very deep bottom, in most instances — an immaterial source and 
explanation; that which we call reality arises in the last analysis from the posing of 
yes — no questions and the registering of equipment-evoked responses; in short, that all 
things physical are information-theoretic in origin and that this is a participatory 
universe.
\end{quotation}

Recent experiments with entangled photons suggests that indeed, an act of observation 
or even a choice of measurement performed on one photon influence the results obtained 
for the other one. In an very recent experiment with photon heralding, the act of observation of 
one photon has brought effectively the entangled twin photon into existence 
\cite{JQI_OL38_1609}. 
Another recent experiment demonstrated that a causally separated choice of measurement can
influence the behaviour of a quantum object, i.e. whether it behaves like a wave or 
like a particle 
\cite{PNAS110_1221}. These results are therefore strong {\em experimental} indications supporting 
the Wheeler idea of a participatory universe, at least in the quantum realm, and in consequence 
also the  ``It from Bit'' idea. 

Another even stronger conclusion is that if we want to live in peace with the special theory of 
relativity, an individual photon cannot be treated sometimes as behaving definitely as a wave and
sometimes behaving definitely as a particle. This leads directly to the thought, that a proper formal 
description of reality cannot be continuous (for waves) or discrete (for particles), but have 
to be in some sense {\em simultaneusly} continuous and discrete. 

According to the Nyquist-Shannon theorem \cite{Nyquist,Shannon}, a fundamental result in information
theory, continuous information may be converted into an equivalent discrete information by sampling,
provided that the sampling rate exceeds the doubled bandlimit frequency. Such an approach to physical
fields and also to the spacetime itself has been suggested awhile ago by Kempf \cite{Kempf1,Kempf2}.

Bits, trits or other commonly used units of information are not very convenient to represent
complicated and structured information, not to mention an ``It from Bit'' programme. 
What is really needed is rather a set of universal ``building blocks'' with an appropriate capacity 
for encoding continuous information,  ``generalized bits'' in a sense. 

An interesting but not widely known theorem by Kolmogorov on superpositions of continuous functions 
\cite{Kolmogorov} could pave the way in this direction and has also very interesting interpretative
implications. Namely, without going into formal details, this theorem states that {\em every} 
multivariate continuous function could be represented in the following simple form:
%%%
\begin{equation} 
f(x_1, \ldots, x_n) = \sum_{i=0}^{2n} g\left( \sum_{j=1}^{n} \lambda_{j}\phi_{i}(x_{j}) \right)  
\end{equation} 
%%%   
where all constants $\lambda_{j}$ and also all (continuous) functions $\phi_{i}(x_{j})$ of one 
variable do {\em not} depend on the function $f$. The function $g$ is in turn a continuous function 
of one variable, remaining in a one-to-one relationship to $f$. 
Therefore $(\lambda_{j}, \phi_{i}(x_{j}))$ are perfect candidates for our ``generalized bits''. 

For almost 30 years this theorem was treated mainly as a fine toy for pure mathematicians, without 
practical use, until the observation was made by Hecht-Nielsen \cite{HechtNielsen1} that Kolmogorov 
theorem could describe neural networks. Moreover, it turned out at the same time, that due to the 
Kolmogorov theorem it is also possible to encode an arbitrary continuous multivariate function in 
a three-layer neural network with continuous activation functions 
(cf. also the book \cite{HechtNielsen2}). Therefore, an universal neural network representation
is also possible, at least in principle.

A strong indication that this is a step in the right direction when searching for ``generalized bits'' 
for an ``It from Bit'' programme is provided by the explanation facilitated by the Kolmogorov theorem, 
why the fundamental physical equations are of second order \cite{YamakawaKreinovich}. Especially 
interesting in this context is the nonsmoothness of fundamental phenomena resulting from the fact 
that smooth multivariate functions cannot be generally represented as superpositions of smooth 
functions of one variable (cf. \cite{Vitushkin} for details).
With this ``inherent unsmoothess'' it is therefore enough room left for all the weird behavior 
observed in Nature.

Are we able to determine the ``generalized bits'' $(\lambda_{j}, \phi_{i}(x_{j}))$ somehow explicitly
or calculate them out ?  At present, we know that they do exist and are computable (cf. \cite{BraunGriebel,Brattka} for recent rigorous results). 
Hence, a (pure) mathematician could probably sleep well with that, but a (genuine) physicist most 
probably not at all. From the proofs presented in the cited papers it is obvious that  
``generalized bits'' a'la Kolmogorov are generally {\em very} weird objects.

In conclusion, the "Bit from It" idea, although strange at the first sight, seems plausible, 
provided we use ``generalized bits''. Another attractive idea, only mentioned in this essay but 
closely related to the ``generalized bits'' approach is the universal neural network representation,
something like the ``Calculating Space'' of Konrad Zuse \cite{Zuse}, arguably the first attempt
to get the {\em It} from {\em Bit}.

\newpage
%%%%%%%%%%%%%%%%%%%%%%%%%%%%%%%%%%%%%%%%%%%%%%%%%%%%%%%%%%%%%%%%%%%%%%%%%%%%%%%%%%

\end{document}